\documentclass[aps,prb,floatfix,epsfig,twocolumn,showpacs,preprintnumbers]{revtex4}
\usepackage{amsmath}
\usepackage{graphicx}
\usepackage{dcolumn}
\usepackage{bm}

\begin{document}

\title{Electron-Phonon Superconductivity in LaO$_{0.5}$F$_{0.5}$BiSe$_{2}$}
\author{Yanqing Feng,$^{1}$ Hang-Chen Ding,$^{2}\ $Yongping Du,$^{1}\ $%
Xiangang Wan,$^{1\ast }$ Bogen Wang,$^{1}\ $Sergey Y. Savrasov$^{3}$ and
Chun-Gang Duan$^{2,4}$}
\affiliation{$^{1}$Department of Physics and National Laboratory of Solid State
Microstructures, Nanjing University, Nanjing 210093, China\\
$^{2}$Key Laboratory of Polar Materials and Devices, Ministry of Education,
East China Normal University, Shanghai 200062, China \\
$^{3}$Department of Physics, University of California, Davis, One Shields
Avenue, Davis, CA 95616, USA \\
$^{4}$National Laboratory for Infrared Physics, Chinese Academy of Sciences,
Shanghai 200083, China}
\date{\today }

\begin{abstract}
We report density functional calculations of the electronic structure, Fermi
surface, phonon spectrum and electron--phonon coupling for newly discovered
superconductor LaO$_{0.5}$F$_{0.5}$BiSe$_{2}$. Significant similarity
between LaO$_{0.5}$F$_{0.5}$BiS$_{2}$ and LaO$_{0.5}$F$_{0.5}$BiSe$_{2}$ is
found, i.e. there is a strong Fermi surface nesting at ($\pi $,$\pi $,0),
which results in unstable phonon branches. Combining the frozen phonon total
energy calculations and an anharmonic oscillator model, we find that the
quantum fluctuation prevents the appearance of static long--range order. The
calculation shows that LaO$_{0.5}$F$_{0.5}$BiSe$_{2}$ is highly anisotropic,
and same as LaO$_{0.5}$F$_{0.5}$BiS$_{2}$, this compound is also a
conventional electron-phonon coupling induced superconductor.
\end{abstract}

\date{\today }
\maketitle

\section{INTRODUCTION}

In 2012, a new superconductor Bi$_{4}$O$_{4}$S$_{3}$ has been found.\cite%
{Bi4O4S3} This material has a layered structure composed of a stacking of
rock-salt-type BiS$_{2}$ layers and Bi$_{4}$O$_{4}$(SO$_{4}$)$_{1-x}$
blocking layers.\cite{Bi4O4S3} The stacking structure of the superconducting
and blocking layer is analogous to those of high-$T_{c}$ cuprates,\cite{HTC}
and iron pnictides.\cite{Fe-based} Thus the discovery of Bi$_{4}$O$_{4}$S$%
_{3}$\ has immediately triggered a wave of extensive studies.\cite%
{Bi4O4S3,LaOFBiS2,NdOBiS2,Bi4O4S3-2,LaO0.5F0.5BiS2,Singh,Bi4O4S3-3,Se-doping,PrO0.5F0.5BiS2,LnOBiS2,LaMOBiS2,SrLaBiS2,s-wave-2,s-wave,lattice vibrations,Wen-HH-2,Wen-HH-3,NdOBiS2-single-crystal,Minimal electronic models,Wan,Yildirim,Xing,SpinExc,Dagoto-RPA,QH-Wang,Hong-Yao,JP-Hu,SL-Liu,SrLaBiS2-2}
In addition to Bi$_{4}$O$_{4}$S$_{3}$, several new BiS$_{2}$-based
superconductors had been synthesized: LnO$_{1-x}$F$_{x}$BiS$_{2}$ (Ln=La,
Nd, Ce, Pr and Yb),\cite%
{LaOFBiS2,NdOBiS2,Bi4O4S3-2,LaO0.5F0.5BiS2,PrO0.5F0.5BiS2,LnOBiS2} Sr$_{1-x}$%
La$_{x}$FBiS$_{2}$\cite{SrLaBiS2,SrLaBiS2-2} and La$_{1-x}$\textit{M}$_{x}$%
OBiS$_{2}$ (\textit{M}=Ti, Zr, Hf and Th).\cite{LaMOBiS2} The common feature
for these compounds is that they all have the same superconducting BiS$_{2}$%
\ layer. Understanding the mediator of pairing as well as the pairing
symmetry for this new layered superconductor is therefore a fundamental
issue, and attracts a lot of research attention.\cite%
{s-wave-2,s-wave,Wen-HH-2,Wen-HH-3,lattice vibrations,Minimal electronic
models,Wan,Yildirim,Xing,SpinExc,Dagoto-RPA,QH-Wang,Hong-Yao,JP-Hu,SL-Liu,NdOBiS2-single-crystal}

Several theoretical works have been reported, especially for LaO$_{0.5}$F$%
_{0.5}$BiS$_{2}$,\cite{LaOFBiS2} the compound that posses the highest $T_{c}$
among known BiS$_{2}$ based materials and whose structure is similar to
superconducting iron arsenides LaFeO$_{1-x}$F$_{x}$As.\cite{Fe-based} It has
been found that the bands crossing Fermi level are Bi-6\textit{p} states and
a two \textit{p} bands electronic model has been proposed based on band
structure calculation.\cite{Minimal electronic models} Due to the
quasi-one-dimensional nature of the conduction bands, a good Fermi--surface
nesting with wave vector $\mathbf{k}$=($\pi ,\pi ,0)$ has been found.\cite%
{Minimal electronic models} The lattice dynamics and electron-phonon
interaction of LaO$_{0.5}$F$_{0.5}$BiS$_{2}$\ have also been studied using
density functional theory based calculations\cite{Wan,Yildirim,Xing}. It has
been suggested that due to the Fermi surface nesting a charge-density-wave
(CDW) instability around $M$\ point is essential.\cite{Wan} An
ferroelectric-like soft phonon mode have also been proposed.\cite{Yildirim}
Basically all the density-functional linear response calculations give a
large electron-phonon coupling constant ($\lambda \sim 0.8$), and suggest LaO%
$_{1-x}$F$_{x}$BiS$_{2}$ as a strong electron-phonon coupled conventional
superconductor.\cite{Wan,Yildirim,Xing}

In contrast to the band structure calculation, there are also works
emphasizing the importance of electron-electron interaction and the
possibility of unconventional superconductivity.\cite%
{Dagoto-RPA,SpinExc,QH-Wang,Hong-Yao,JP-Hu} Starting from the two-orbital
model, the spin/charge fluctuation mediated pairing interactions had been
studied by using random-phase approximation,\cite{Dagoto-RPA,SpinExc} and an
extend \textit{s}-wave or \textit{d}-wave pairing had been proposed.\cite%
{Dagoto-RPA} With the assumption that the pairing is rather short range
interaction, Liang \textit{et al.} find that the extended \textit{s}-wave
pairing symmetry is very robust.\cite{JP-Hu} Possible triplet pairing and
weak topological superconductivity had been suggested based on
renormalization-group numerical calculation.\cite{QH-Wang} It had also been
proposed that BiS$_{2}$ based superconductor possess type-II two-dimensional
Van Hove singularities, and the logarithmically divergent density of states
may induce unconventional superconductivity.\cite{Hong-Yao}

There are also debates about the pairing symmetry experimentally. The
temperature dependence of magnetic penetration depth have been measured by
tunnel diode oscillator technique, and it had been suggested that BiS$_{2}$
layered superconductors are conventional \textit{s}-wave type superconductor
with fully developed gap.\cite{s-wave-2} Muon-spin spectroscopy measurements
($\mu $SR) shows a marked two-dimensional character with a dominant \textit{s%
}-wave temperature behavior.\cite{s-wave} On the other hand, both the
experimental upper critical field, which exceeds the Pauli limit, and the
large ratio $2\Delta /T_{c}\sim 16.6$ imply that the superconductivity is
unconventional.\cite{Wen-HH-2} Recently, NdO$_{1-x}$F$_{x}$Bi$_{1-y}$S$_{2}$
single crystals had been grown.\cite{NdOBiS2-single-crystal,Wen-HH-3}
Resistivity and magnetic measurements reveal that the superconductivity is
really derived from the materials intrinsically.\cite{Wen-HH-3} Moreover, a
giant superconducting fluctuation and anomalous semiconducting normal state
have been found for the single crystal sample, suggesting that the
superconductivity in this newly discovered superconductor may not be
formatted into the BCS theory\cite{Wen-HH-3}.

Very recently, a new superconductor LaO$_{0.5}$F$_{0.5}$BiSe$_{2}$ had been
discovered.\cite{BiSe2} LaO$_{0.5}$F$_{0.5}$BiSe$_{2}$ has similar structure
of LaO$_{0.5}$F$_{0.5}$BiS$_{2}$ yet with a lower T$_{c}\ $($\sim $2.6 K).%
\cite{BiSe2} In order to shed light on the superconducting nature of this
family, it is essential to investigate LaO$_{0.5}$F$_{0.5}$BiSe$_{2}$. Here
we report our theoretical studies of the electronic structure and lattice
dynamic properties for LaO$_{0.5}$F$_{0.5}$BiSe$_{2}$. Our first-principles
calculation shows that the band structure of LaO$_{0.5}$F$_{0.5}$BiSe$_{2}$
is quite similar as that of LaO$_{0.5}$F$_{0.5}$BiS$_{2}$, there is also a
strong Fermi surface nesting at $k=(\pi ,\pi ,0)$ (i.e. $M$ point), which
leads to imaginary harmonic phonons at this $k$ point associated with
in-plane displacements of Se atoms. Although the $\sqrt{2}\times \sqrt{2}%
\times 1$\ supercell frozen phonon calculations confirm a double well
related to the soft-mode located at $M$ point, our effective model reveals
that the double well is too shallow, consequently the CDW structural phase
transition indeed will not happen. Replacing S by Se will reduce the Deybe
frequency $\omega _{D}$\ as well as the electron-phonon coupling constant $%
\lambda $, which explains the decrease of $T_{c}$.

\section{COMPUTATIONAL METHOD}

Our electronic structure calculations are performed based on the Quantum
ESPRESSO package (QE)\cite{pwscf} and ultrasoft pseudopotentials\cite{the
ultrasoft pseudopotential} and the generalized gradient approximation of
Perdew, Burke, and Ernzerhof (PBE).\cite{GGA} The basis set cutoff for the
wave functions was 60 Ry while 600 Ry cutoff was used for the charge
density. A dense 18$\times $ 18$\times $ 6 \emph{k}--point mesh had been
used in the irreducible Brillouin zone (IBZ) for self--consistent
calculations. For structural optimization, the positions of ions were
relaxed towards equilibrium until the Hellman--Feynman forces became less
than 2 meV/\AA . Same as LaO$_{0.5}$F$_{0.5}$BiS$_{2}$,\cite{Wan,Yildirim}
we find that for LaO$_{0.5}$F$_{0.5}$BiSe$_{2}$, spin-orbital coupling (SOC)
plays only a marginal role on the electronic states near Fermi level ($E_{F}$%
)\ and lattice dynamic properties. Thus we neglect it and adopt the scalar
relativistic version of QE.\cite{pwscf}

\section{RESULTS AND DISCUSSIONS}

\begin{table}[tbp]
\caption{The calculated lattice parameters and Wyckoff positions of LaO$%
_{0.5}$F$_{0.5}$BiSe$_{2}$. Experimental results are also listed for
comparison\protect\cite{BiSe2}.}%
\begin{tabular}{lrrr}
\hline
& Site & Cal. & Exp. \\ \hline
\emph{a}({\AA }) &  & {4.1594} & {4.1753} \\
\emph{c}({\AA }) &  & {14.0156} & {14.2634} \\
\emph{z} & La (2\emph{c}) & {0.0943} & {\ 0.0947} \\
\emph{z} & Bi (2\emph{c}) & {0.6204} & {\ 0.6136} \\
\emph{z} & Se1 (2\emph{c}) & {0.3847} & {\ 0.3933} \\
\emph{z} & Se2 (2\emph{c}) & {0.8115} & {\ 0.8155} \\ \hline
\end{tabular}%
\end{table}

LaO$_{0.5}$F$_{0.5}$BiSe$_{2}$ has a layered crystal structure with a space
group P4/nmm\cite{BiSe2}. This material is formed by alternatively stacking
of BiSe$_{2}$ layers and the blocking layer Bi$_{4}$O$_{4}$(SO$_{4}$)$_{1-x}$%
. La, Se and Bi locate at 2\emph{c} position, while O/F take the 2\emph{a}
site. Being embedded into LaO plane, it had been found that the substitution
O\ by F has only small effect on the BiS$_{2}$ layer in LaO$_{0.5}$F$_{0.5}$%
BiS$_{2}$\cite{Wan}, and the main influence of F\ substitution is a carrier
doping\cite{Wan}. Thus we simulate LaO$_{0.5}$F$_{0.5}$BiSe$_{2}$ by
replacing half of the Oxygen 2\emph{a}-sites by F\ orderly, despite the
substitution may be random in reality. We perform the full structural
optimization including the lattice parameters and atomic positions, the
optimized lattice parameters and Wyckoff positions are shown in Table I,
together with available experimental data \cite{BiSe2}. Our numerical
lattice parameters and internal coordinates are in good agreement with the
experiment as shown in Table I. For LaO$_{0.5}$Bi$_{0.5}$Se$_{2}$, Bi and
Se1 form a nearly perfect plane, while for LaO$_{0.5}$Bi$_{0.5}$S$_{2}$, the
experimental and theoretical height of S1 has a large difference\cite%
{Wan,Yildirim}, and the importance of buckling of S-atoms have been
discussed.\cite{Yildirim}

Based on theoretical lattice structure, we perform band structure
calculation, and find that the electronic structure of LaO$_{0.5}$Bi$_{0.5}$%
Se$_{2}$ is similar with that of LaO$_{0.5}$Bi$_{0.5}$S$_{2}.$\cite{Minimal
electronic models,Wan,Yildirim} As shown in Fig.1, the dispersion along $%
\Gamma $\ to $Z$ line is quite small, clearly implying a two dimensional
character of the band structure and the negligible interlayer hybridization.
La states, which appear considerably above the $E_{F}$, has almost no
contribution around $E_{F}$. O/F 2\textit{p} states are almost fully
occupied and mainly located between -5.0 and -2.0 eV. Hybridized with Bi 6%
\textit{p} states, Se2 4\textit{p} states have wider bandwidth, but this
state is also located primarily below the $E_{F}$, has negligible
contribution around the $E_{F}$. Bi and Se1 form a layer, consequently these
states have strong hybridization. The bands around the $E_{F}$ basically
come from Bi 6\textit{p} with also small contribution from Se1 4\textit{p }%
as shown in Fig.1. The density of state at $E_{F}$ is equal to N($E_{F}$%
)=1.97 eV$^{-1}$ per unit cell. This corresponds to a bare Sommerfeld
specific heat coefficient $\gamma _{bare}$=2.42 mJ$/$mol K$^{2}$, which is
just slightly smaller than the numerical value ($\sim $3.0 mJ$/$mol K$^{2}$)
of LaO$_{0.5}$Bi$_{0.5}$S$_{2}$.\cite{Xing} The calculated bare plasma
frequencies are $\hbar \omega _{p,xx}=\hbar \omega _{p,yy}=5.99\ eV$\ and $%
\hbar \omega _{p,zz}=0.12\ eV$, which corresponds to a very large anisotropy
$\sigma _{xx}/\sigma _{zz}\sim 2500$\ in the assumption of constant
scattering time. This huge anisotropy may be detected via optical or
transport measurement for single crystal sample.
\begin{figure}[tbp]
\includegraphics[width=3.5in]{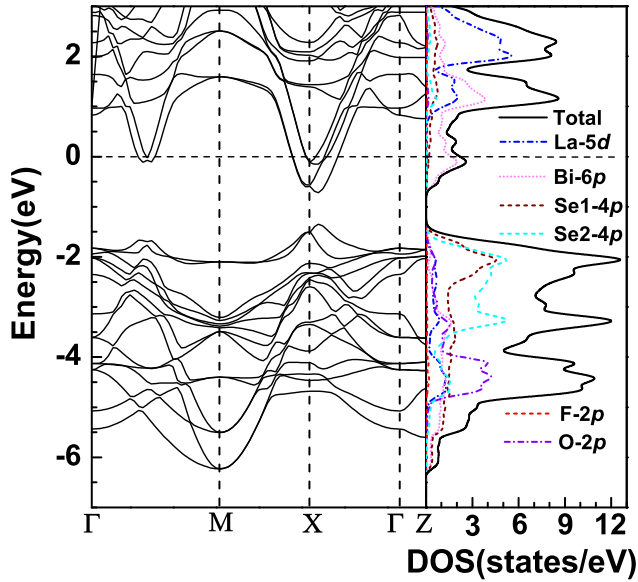}
\caption{(Color online) Calculated band structure and density of state of LaO%
$_{0.5}$F$_{0.5}$BiSe$_{2}$. }
\end{figure}

\begin{figure}[tbp]
\includegraphics[width=3.5in]{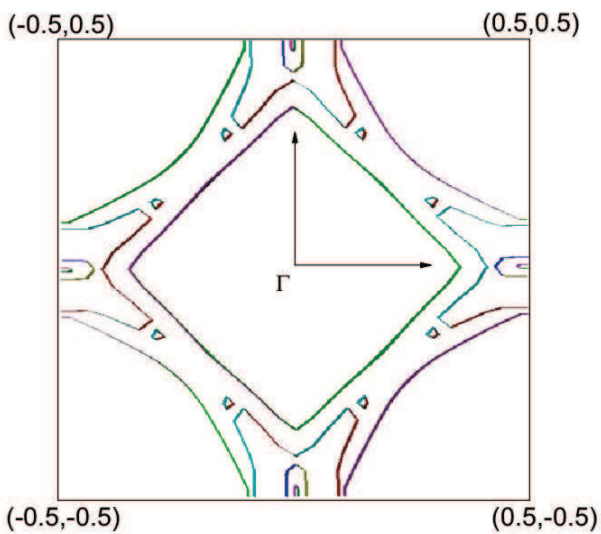}
\caption{(Color online) Calculated Fermi surface of LaO$_{0.5}$F$_{0.5}$BiSe$%
_{2}$: cross section for k$_{z}$=0.}
\label{Fig-FS}
\end{figure}

We also calculate the Fermi surface and show the results in Fig.2. Very
similar to LaO$_{0.5}$F$_{0.5}$BiS$_{2}$,\cite{Minimal electronic models,Wan}
the Fermi surface of LaO$_{0.5}$F$_{0.5}$BiSe$_{2}$ are also two
dimensional-like and there is a Fermi surface nesting at wavevector near $%
\mathbf{k}=(\pi ,\pi ,0)$.

Density functional linear response approach has been proven to be very
successful in the past to describe electron--phonon interactions and
superconductivity in metals \cite{EPI}, including its applications to
Plutonium, \cite{Pu} MgB$_{2}$, \cite{MgB2} and many other systems. Here we
apply this first-principles linear response phonon calculation \cite%
{Linear-response} as implemented in QE\cite{pwscf} to study LaO$_{0.5}$F$%
_{0.5}$BiSe$_{2}$. An 18$\times $18$\times $6 grid was\ used for the
integration over the IBZ. We show the calculated phonon spectrum along major
high symmetry lines of the IBZ in Fig.3. Same as LaO$_{0.5}$F$_{0.5}$BiS$%
_{2},$\cite{Wan,Yildirim,Xing} the phonon modes have only a little
dispersion along $\Gamma $-Z direction, which again indicates the smallness
of the interlayer coupling. The phonon dispersions are extend up to 350 cm$%
^{-1}$, which is smaller than that of LaO$_{0.5}$F$_{0.5}$BiS$_{2}$\cite%
{Wan,Yildirim}.\ There are basically two panels in the phonon spectrum. The
top six branches above 180 cm$^{-1}$ are almost completely contributed by O
and F, while the Bi-Se vibration dominate in the low frequency region.
Comparing with LaO$_{0.5}$F$_{0.5}$BiS$_{2},$\cite{Wan,Yildirim,Xing}, LaO$%
_{0.5}$F$_{0.5}$BiSe$_{2}$ has more branches locate at low frequency region.
Analyzing the evolution of the phonon eigenvectors in the IBZ reveals that
there is clear separation between the \textit{xy} and \textit{z} polarized
vibrations, and the phonon modes with large dispersion mainly comes from the
Bi-Se1 in-plane vibration.

\begin{figure}[tbp]
\includegraphics[width=4.0in]{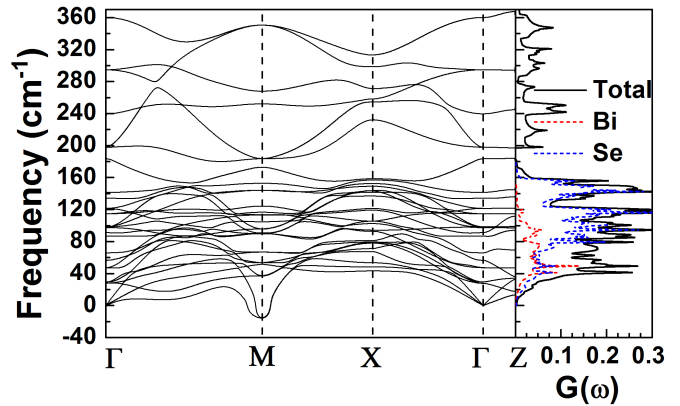}
\caption{(Color online) Calculated phonon dispersions and phonon density of
state of LaO$_{0.5}$F$_{0.5}$BiSe$_{2}$.}
\end{figure}

There are also unstable modes located around $M$ point as shown in Fig.3. We
associate it with the strong Fermi surface nesting. The number of soft modes
in LaO$_{0.5}$F$_{0.5}$BiSe$_{2}$\ is two, while LaO$_{0.5}$F$_{0.5}$BiS$%
_{2} $\ has four unstable modes around $M$ point\cite{Wan,Yildirim}. From
the analysis of the calculated polarization vectors,\ we find that these two
unstable modes are mainly contributed by the Se1 in-plane vibrations.

\begin{figure}[tbp]
\includegraphics[width=3.5in]{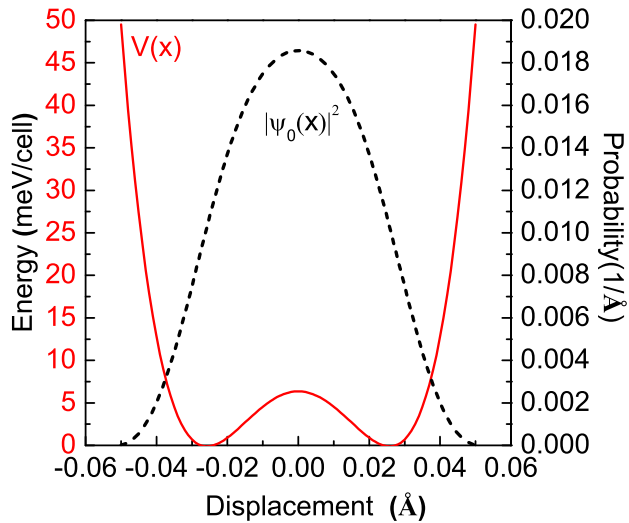}
\caption{(Color online) Calculated double well potential (red line) for the
unstable phonon mode using the frozen phonon method. The probability plot
(black dot line) of the ground state atomic wave function is also shown.}
\end{figure}

We then perform a frozen phonon calculation by using a $\sqrt{2}\times \sqrt{%
2}\times 1$ supercell with respect to its original unit cell to adapt the
lattice distortions due to the possible CDW instability associated with the
soft phonon mode located at $M$ point. The atomic motions in the frozen
phonon calculation is chosen according to the eigenvectors of the unstable
phonon modes at the $M$ point, which basically is Se1 in-plane displacement.
The results of these calculations reveal essentially anharmonic interatomic
potentials, and a shallow double well potential ($\sim -6\ meV$ per unit
cell) where the Se1 atoms shift about 0.03 \AA\ away from the original high
symmetry position as show in Fig.4. The depth of the double well is less
than half of that in LaO$_{0.5}$F$_{0.5}$BiS$_{2}.$\cite{Wan} For LaO$_{0.5}$%
F$_{0.5}$BiS$_{2}$\ it had been found that the displacements of S atom are
dynamic.\cite{Wan,Yildirim} To check if the quantum zero-point motions also
prevent the structural distortion of LaO$_{0.5}$F$_{0.5}$BiSe$_{2}$, we
therefore extend the equilibrium position analysis by solving numerically
Schr\"{o}dinger's equation for the anharmonic potential well found from
frozen-phonon calculations, as shown in the red line of Fig. 4. Indeed, our
numerical atomic ground-state wave function is centered at the high symmetry
position, as demonstrated by the probability curve shown in black dot line
of Fig. 4. It is therefore clear that the Se1 displacement is dynamic, and
the unstable phonon modes at $M$ point are not related to a statically
distorted structure of LaO$_{0.5}$F$_{0.5}$BiSe$_{2}$.\ Experimentally the
resistivity changes smoothly from 300 K to about 3 K, and there is not
abnormal behavior due to the gap opening associated with CDW phase transition%
\cite{BiSe2}. This experimental observation is consistent with our above
model calculation.

Finally, we check the linewidth $\lambda _{\nu }(\mathbf{q})$ of all stable
phonons to see the contribution from different modes. Calculation shows that
the O/F modes have negligible contribution to electron-phonon coupling. With
strong hybridization, the coupling, however, is relatively strong for the
BiSe based modes. Counting the contribution only from the stable modes
results in a coupling constant ($\lambda =0.47$) calculated using 4$\times $4%
$\times $2\ q-mesh.

To find the contribution from the anharmonic modes, we follow the expression
introduced by Hui and Allen which generalizes zero-temperature
electron-phonon coupling to the anharmonic case. \cite{Hui} We make an
essential approximation by assuming that the unstable mode is not coupled to
the other modes. The phonon-phonon interactions and finite-temperature
effects are also neglected in this treatment. Basically we need all phonon
excited states.\cite{Hui,Meregalli} Fortunately, it had been found the
convergency in the sum over the virtual phonon states is fast.\cite%
{Hui,Meregalli} By taking the harmonic dipole matrix elements, and
interpreting the first excitation energy from solving the Schr\"{o}dinger's
equation for the anharmonic atomic\ potential well as the phonon energy,\cite%
{Hui} we estimate the electron-phonon coupling for the two anharmonic modes
at the $M$ point. Our numerical $\lambda $\ value from this two modes is
about 0.04. Adding this value with the contribution from all stable modes
gives us a total coupling constant of 0.51, which is about half of that of
LaO$_{0.5}$F$_{0.5}$BiS$_{2}.$\cite{Wan} The estimated Deybe temperature ($%
\omega _{D}=220\ K$) is also smaller than that of LaO$_{0.5}$F$_{0.5}$BiS$%
_{2}.$\cite{Wan} With these values and taking the Coulomb parameter $\mu
^{\ast }\simeq 0.1$, McMillan formula yields values of $T_{c}\simeq 2.40\ K$
in reasonable agreement with the experiment.\cite{BiSe2}

\section{CONCLUSIONS}

We have studied the electronic structure, lattice dynamics and
electron--phonon interaction of the newly found superconductor LaO$_{0.5}$F$%
_{0.5}$BiSe$_{2}$ using density functional theory and linear response
approach. Same as in the case of its cousin LaO$_{0.5}$F$_{0.5}$BiS$_{2}$, a
strong Fermi surface nesting is also found at ($\pi $,$\pi $,0), which
results in phonon softening, and we find that the quantum fluctuation
prevents the appearance of static long-range order. Considering both
harmonic and anharmonic contributions to electron-phonon coupling, we obtain
a coupling constant $\lambda \sim 0.51$, which is capable of producing the
experimental $T_{c}$\ value and suggesting LaO$_{0.5}$F$_{0.5}$BiSe$_{2}$ as
a electron--phonon superconductor.

\acknowledgements X.G.W acknowledges useful conversations with Prof. H.H.
Wen, J.X. Li, Q.H. Wang, Donglai Feng and Dawei Shen. This work was
supported by the National Key Project for Basic Research of China (Grants
No. 2011CB922101, 2010CB923404, 2013CB922301 and 2014CB921104), NSFC under
Grants No. 91122035, 11174124, 11374137, 61125403), PAPD, Program of
Shanghai Subject Chief Scientist, PCSIRT. Computations were performed at the
ECNU computing center.

$^{\ast }$xgwan@nju.edu.cn

\end{document}